\documentclass[reprint,amsmath,amssymb,aps,pra,groupedaddress,nofootinbib,twocolumn,superscriptaddress]{revtex4-1}
\usepackage[margin=1.8cm]{geometry}
\usepackage{graphicx}
\usepackage{amsthm,amssymb,amsmath,braket,mathdots,mathtools}
\usepackage{bm}
\usepackage[colorlinks,bookmarks=true,citecolor=blue,linkcolor=red,urlcolor=blue]{hyperref}
\usepackage{epstopdf,psfrag}
\usepackage{relsize,amsbsy}
\usepackage[export]{adjustbox}

\usepackage{dsfont}

\usepackage{graphicx,xcolor,tikz}
\usepackage{layouts}

\usepackage{graphicx}
\usepackage{ragged2e}

\newcommand{\1}{\mathds{1}}

\newcommand{\be}{\begin{equation}}
\newcommand{\ee}{\end{equation}}

\newcommand{\bit}{\begin{enumerate}}
	\newcommand{\eit}{\end{enumerate}}

\definecolor{bananayellow}{rgb}{1.0, 0.88, 0.21}
\definecolor{straw}{rgb}{0.32, 0.28, 0.1}

\begin{document}

\title{Disorder-free localization in a simple $U(1)$ lattice gauge theory}
\author{Irene Papaefstathiou}
\affiliation{Max-Planck-Institut f{\"u}r Quantenoptik, Hans-Kopfermann-Str. 1, D-85748 Garching, Germany}
\affiliation{Blackett Laboratory, Imperial College London, London SW7 2AZ, United Kingdom}
\author{Adam Smith}
\affiliation{Department of Physics CMT, Technische Universit{\"a}t M{\"u}nchen, James-Franck-Stra{\ss}e 1, D-85748 Garching, Germany}
\author{Johannes Knolle}
\affiliation{Department of Physics CMT, Technische Universit{\"a}t M{\"u}nchen, James-Franck-Stra{\ss}e 1, D-85748 Garching, Germany}
\affiliation{Munich Center for Quantum Science and Technology (MCQST), 80799 Munich, Germany}
\affiliation{Blackett Laboratory, Imperial College London, London SW7 2AZ, United Kingdom}

\date{\today}

\begin{abstract}
Localization due to the presence of disorder has proven crucial for our current understanding of relaxation in isolated quantum systems. The many-body localized phase constitutes a robust alternative to the thermalization of complex interacting systems, but recently the importance of disorder has been brought into question. A number of disorder-free localization mechanisms have been put forward connected to local symmetries of lattice gauge theories. Here, starting from translationally invariant $(1+1)$-dimensional quantum electrodynamics, we modify the dynamics of the gauge field which allows us to construct a lattice model with a U(1) local gauge symmetry revealing a mechanism of disorder-free localization. We consider two different discretizations of the continuum model resulting in a free-fermion soluble model in one case and an interacting model in the other. We diagnose the localization of our translationally invariant model in the far-from-equilibrium dynamics following a global quantum quench.
\end{abstract}

\maketitle

\section{Introduction} 
Revealing the effect of disorder has been crucial for our understanding of how complex quantum systems can relax. It was found by Anderson~\cite{anderson1958absence} that the presence of disorder leads to a perfect insulator in non-interacting systems due to the exponential localization in space of single particle wavefunctions. The diagnostics of Anderson localization include the non-decaying return probability of a state that was localized initially and the localization of the eigenstates of the Hamiltonian. Localization can also be diagnosed through the spectral properties of the system ~\cite{scardicchio2017perturbation}. Importantly, in interacting systems it is possible to have a robust many-body localized phase (MBL) of matter~\cite{basko2006metal,Znidaric2008, Bardason2012,abanin2017,nandkishore2015many}. Unlike integrable or Anderson localized systems, MBL systems are stable to generic perturbation and are not protected by symmetry or local conservation laws. MBL provides an alternative to the more generic expectation that complex interacting systems should be ergodic and eventually thermalize~\cite{DAlessio2016,Deutsch1991,Srednicki1994}.

More recently, the question has been posed whether disorder is a requirement for localization or whether this non-ergodic behaviour is possible in translationally invariant systems. An alternative was demonstrated in a model related to a $\mathbb{Z}_2$ lattice gauge theory (LGT)~\cite{smith2017disorder}, where the disorder-free localization mechanism emerged from the local gauge symmetry. This mechanism has since been utilised in a range of simple $\mathbb{Z}_2$ LGTs~\cite{prosko2017}, the massless one-dimensional lattice Schwinger model of quantum electrodynamics (QED)~\cite{brenes2018many}, and a two-dimensional $U(1)$ quantum link model~\cite{Karpov2020}. As well as the LGT approach, localization without disorder has also been achieved using a variation of Wannier-Stark localization~\cite{vanNieuwenburg9269, stark} and there is also evidence in some kinetically constrained models~\cite{Horssen2015,Hickey2016}.

Disorder-free localization in LGTs is remarkable as they are central in the study of interacting physics, from the description of the standard model of particle physics to strongly-correlated phases of many-body systems~\cite{Lee2006doping}. Even the simplest $(1+1)$-dimensional $U(1)$ Schwinger model of quantum electrodynamics (QED)~\cite{schwinger1962} captures the Schwinger mechanism of particle-antiparticle pair production~\cite{schwinger1951} and is a toy model of quark confinement~\cite{hamer1982massive}. The recent observation that LGTs can form the basis for disorder-free localization mechanisms~\cite{smith2017disorder,brenes2018many} has led to a surge of interest from the condensed matter perspective. In this paper we study a model, which is the $U(1)$ analogue of the $\mathbb{Z}_2$ LGT of Ref.~\cite{smith2017disorder} whose exact solubility has enabled rigorous results for its localization properties. The lattice model studied here with a U(1) gauge symmetry is constructed from a modification of the massless one-dimensional lattice Schwinger model of QED, using the Kogut-Susskind formulation of Staggered fermions~\cite{kogut1975j}. The latter is manifestly a model of long-range spin-spin interactions.

The lattice discretization of a relativistic continuum field theory is not unique and is complicated by the Nielsen-Ninomiya no-go theorem~\cite{nielsen1981absence} and the fermion doubling problem~\cite{montvay1997quantum,karsten1981lattice}. The former states that in any discretization, one of the following symmetries must be broken: hermiticity, locality, chiral symmetry or discrete translation invariance. The fermion doubling problem refers to non-physical degrees of freedom that result from the naive discretization. Out of the several possible approaches, in this paper we consider Staggered fermions and Wilson fermions~\cite{rothe2005lattice}. Staggered fermions break the discrete translational symmetry whereas Wilson fermions do not respect chiral symmetry.

We start with the Schwinger model in the continuum limit and we modify the dynamics of the gauge field, with the aim of obtaining a soluble model with a disorder-free mechanism for localization. We first consider the Staggered fermions discretization, which leads to a model of free-fermions with an emergent disorder-free localization mechanism. This solubility allows us to identify the localized behaviour and perform large scale numerical simulations. In addition, we diagnose the localization using the persistence of local information under far-from-equilibrium dynamics after a global quantum quench~\cite{Smith2018,Hauschild2016}. These quench protocols are relevant to experiments, such as those in cold atom optical lattice~\cite{schneider2012fermionic, Choi1547}, where they are routinely used to diagnose localization behaviour. Interestingly, in contrast to Staggered fermions, the Wilson fermions discretization of the model gives rise to fermionic interactions. 


This paper is structured as follows. We first review in Sec.~\ref{sec:Z2LGT} and in Sec.~\ref{sec: Schwinger model} the two models that motivate our model of interest. We begin by introducing the $\mathbb{Z}_2$ LGT from Ref.~\cite{smith2017disorder} in Sec.~\ref{sec:Z2LGT} and in Sec.~\ref{sec: Schwinger model} we introduce the Schwinger model on the lattice with Staggered fermions and highlight that it is mapped to a model of long-range spin-spin interactions. We then proceed with presenting the Schwinger model in the continuum. With the aim of finding a model of free fermions on the lattice, in Sec.~\ref{sec: continuum} we define a modified continuum $U(1)$ model and we highlight the difference from the standard (1+1)-dimensional QED. We provide the details of the Staggered fermion discretization of this model of interest resulting in a one-dimensional lattice gauge theory. We then study the model, reveal the disorder-free mechanism for localization and demonstrate this behaviour using global quench protocols in Sec.~\ref{sec: disorder}. Following this, in Sec.~\ref{sec: wilson} we consider the Wilson fermion approach for the lattice discretization. Finally, we close with a discussion of our results and an outlook.

\section{The $\mathbb{Z}_2$ LGT}\label{sec:Z2LGT}

We begin by reviewing the model for disorder-free localization introduced in Ref.~\cite{smith2017disorder}. This model is a one-dimensional $\mathbb{Z}_2$ LGT consisting of spinless fermions $\hat{f}_{i}$ that are coupled to spins $\hat{\sigma}_{i,i+1}$ with the translationally invariant Hamiltonian
\begin{equation}\label{eq: Z2Hamiltonian initial}
    \begin{aligned}
    \hat{\mathcal{H}}_{\mathbb{Z}_2}=-J\sum_{\langle i j \rangle}\hat{\sigma}^{z}_{i,j}\hat{f}^{\dagger}_{i}\hat{f}_{j}-h\sum_{i}\hat{\sigma}^{x}_{i-1,i}\hat{\sigma}^{x}_{i,i+1},
   \end{aligned}
\end{equation}
where $J$ and $h$ are constants that represent the tunneling strength and the Ising coupling correspondingly. 
It can be shown that the model holds a set of local conserved quantities $\{q_{j}\}$ which arise from gauge invariance.
By using a duality mapping
\begin{equation}
    \begin{gathered}
        \hat{\tau}^{z}_{j}=\hat{\sigma}^{x}_{j-1,j}\hat{\sigma}^{x}_{j,j+1},\qquad \hat{\sigma}^{z}_{j,j+1}=\hat{\tau}^{x}_{j}\hat{\tau}^{x}_{j+1},
    \end{gathered}
\end{equation}
the Hamiltonian in a given charge sector then assumes the following form
\begin{equation}\label{eq: Z2Hamiltonian}
    \begin{aligned}
    \hat{\mathcal{H}}_{\{q_{j}\}}=-J \sum_{\langle ij\rangle}\hat{c}^{\dagger}_{i}\hat{c}_{j}+2h\sum_{j}q_{j}(\hat{c}^{\dagger}_{j}\hat{c}_{j}-1/2),
   \end{aligned}
\end{equation}
where $\{q_{j}\}$ is a particular set of local conserved quantities with values $\pm1$ on each lattice site and $\hat{c}_{j}=\hat{\tau}^{x}_{j}\hat{f}_{j}$. Eq.~\eqref{eq: Z2Hamiltonian} is a tight-binding model with a disorder potential controlled by the chosen set of conserved charges $\{q_{j}\}$.


\section{The Schwinger model}\label{sec: Schwinger model} 

The goal of this work is to generalise the $\mathbb{Z}_2$ LGT to a $U(1)$ LGT which is soluble and also has an emergent disorder-free mechanism for localization. We do so in analogy with the discretization of the $U(1)$ Schwinger model for QED. Let us therefore briefly review the Schwinger model on the lattice with Staggered fermions and how this is derived from a continuum model. 

The lattice Schwinger model consists of spinless (single-component) fermion operators, $\hat{\Phi}_n$, where $n$ are site indices for a one-dimensional chain with lattice spacing $a$, coupled to a gauge field. The corresponding discrete Hamiltonian is
\begin{equation}\label{eq: Schwinger lattice}
    \begin{aligned}
    \hat{H} = &\frac{g^{2}a}{2}\sum_{n}\hat{L}_{n}^{2}+ m\sum_{n}(-1)^{n}\hat{\Phi}^{\dagger}_{n}\hat{\Phi}_{n},\\
    &-\frac{i}{2a} \sum_{n}\Big(\hat{\Phi}_{n}^{\dagger}e^{i\hat{\theta}_{n}}\hat{\Phi}_{n+1}-h.c.\Big),
    \end{aligned}
\end{equation}
where the fermion operators satisfy the canonical anti-commutation relations $\{\hat{\Phi}^{\dagger}_{n}, \hat{\Phi}_{m} \}=\delta_{nm}\hat{\mathds{1}}$, and
$\{\hat{\Phi}_{n}, \hat{\Phi}_{m} \}=0$. 

The gauge field operators $\hat{\theta}_{n}$ and $\hat{L}_{n}$ are related to the vector potential field $\hat{A}_{n}^{1}$ and the electric field $\hat{E}_{n}$ correspondingly. In particular, they satisfy
\begin{equation}\label{eq: Quantum Model}
\begin{aligned}
\hat{L}_n &= \hat{E}_n/g,\\
\hat{\theta}_n&= -ag\hat{A}^1_n  ,
\end{aligned}
\end{equation}
where the operators $\hat{\theta}_n$ and $\hat{L}_n$ are defined on the `link' between the two lattice sites $n$ and $n+1$, and have the commutation relations $[\hat{\theta}_n, \hat{L}_m] = i \delta_{nm} \hat{\mathds{1}}$. In terms of these operators the $U(1)$ parallel transporters are $\hat{U}_n = e^{i\hat{\theta}_n}$. The operators $\hat{L}_n$ are generally quantized with spectrum, $L_n = 0, \pm1, \pm2, \pm3...$.

The above Hamiltonian is the lattice Schwinger model using the Kogut-Susskind formulation~\cite{kogut1975j} of Staggered fermions~\cite{susskind1977lattice}. The first term represents the energy stored by the electric field, the second term is the staggered mass term and the last term represents the coupling between the fermionic fields and the gauge fields. By performing a Jordan-Wigner transformation and by eliminating the gauge fields from the above Hamiltonian, it can be shown that the system is mapped to a system of spins with long-range interactions~\cite{Hamer_1997}. These interactions arise from the first term of the Hamiltonian~\eqref{eq: Schwinger lattice}, the term that corresponds to the energy stored by the electric field.
 
 The lattice Schwinger model presented above is a lattice discretization of the Hamiltonian of QED in (1+1) dimensions. Following the convention in high-energy physics, we label the time-dimension $0$ and the space dimension $1$ and work in the temporal gauge. The Hamiltonian of the Schwinger model in the continuum reads~\cite{montvay1997quantum,peskin1995introduction}
 \begin{equation}\label{eq: Schwingercont}
    \begin{aligned}
    \hat{\mathcal{H}}= &-\int dx \Big(i\bar{\Psi}(x)\gamma^{1}\big[\partial_{1}+ig\hat{A}^{1}(x)\big]\Psi(x)\Big)\\
    &+\int dx \Big( m\bar{\Psi}(x)\Psi(x)+\frac{1}{2}\hat{E}^2\Big),
   \end{aligned}
\end{equation}
where  $\partial_{1}$ is the partial derivative with respect to $x$, $\gamma^{0}$ and $\gamma^{1}$ are the gamma matrices in two spacetime dimensions, and $\bar{\Psi}=\Psi^{\dagger}\gamma^{0}$. The Dirac field $\Psi (x)$ has two components. The parameters $m$ and $g$ represent the fermion mass and the coupling constant, respectively.

The first term represents the coupling of the fermionic matter field with the vector gauge field $\hat{A}^{1}(x)$, and the second term is the fermionic mass term. The last term is the energy of the electric field $\hat{E}(x)$, which gives dynamics to the gauge field. Note that the above Hamiltonian is in the temporal gauge, where $A^{0}=0$ and $\hat{E}(x)=-\partial_{0}\hat{A}^{1}(x)$. As mentioned above, the term that corresponds to the energy stored by the electric field, $\int dx \frac{1}{2}\hat{E}^2$, gives rise to long-range spin-spin interactions in the discretized model.

\section{Our Model: Staggered Fermions}\label{sec: continuum} 

With the aim of introducing a soluble $U(1)$ LGT, we start with a continuum model. This has a similar form to the Schwinger Hamiltonian of Eq.~\eqref{eq: Schwingercont}. We then discretize this model using either staggered fermions or Wilson fermions.

Let us start with a (1+1)-dimensional continuum model with a local $U(1)$ gauge symmetry. The Hamiltonian reads
\begin{equation}\label{1}
    \begin{aligned}
    \hat{\mathcal{H}}= &-\int dx \Big(i\bar{\Psi}(x)\gamma^{1}\big[\partial_{1}+ig\hat{A}^{1}(x)\big]\Psi(x)\Big)\\
    &+\int dx \Big( m\bar{\Psi}(x)\Psi(x)+B\big[\partial_{1} \hat{E}(x)\big]^2\Big),
   \end{aligned}
\end{equation}
where we use the same notation as the notation used in the previous section. Here $B$ is a constant with dimensions (length)$^{2}$. Our Hamiltonian differs from QED in $(1+1)$-dimensions~\cite{montvay1997quantum,peskin1995introduction} only in the term that corresponds to the dynamics of the electric field, where we have made the change $\int dx \hat{E}^{2}(x)\rightarrow \int dx[{\partial_{1} \hat{E}(x)}]^2$. We highlight here that we will not study the properties of this continuum model but we write it down as a guidance for our lattice discretization.

If we proceed to naively discretize the Dirac Hamiltonian on the lattice, we encounter the so-called doubling problem resulting in non-physical degrees of freedom, called fermion doublers~\cite{karsten1981lattice}. We can avoid this problem with the use of Staggered~\cite{susskind1977lattice} or Wilson~\cite{wilson1974confinement, wilson1977new} fermions. Both discretizations result in lattice models with a local gauge invariance. We proceed with both discretizations, with the aim of finding a model of free fermions on the lattice. 

We first consider the approach of Staggered fermions, mapping the continuous Dirac fermion field onto the single-component fermion operators, $\hat{\Phi}_n$, with $n$ the site indices
\begin{equation}\label{eq: quantum link H}
    \begin{aligned}
    \hat{H}=&\frac{B}{a}\sum_{n}\big(\hat{E}_{n}-\hat{E}_{n-1}\big)^{2}\\
    &-\frac{i}{2a}\sum_{n}\Big( \hat{\Phi}_{n}^{\dagger}e^{-iag\hat{A}_{n}^{1}}\hat{\Phi}_{n+1}-h.c.\Big)\\
    &+m\sum_{n}(-1)^{n}\hat{\Phi}^{\dagger}_{n}\hat{\Phi}_{n},
     \end{aligned}
\end{equation}
where the fermion operators satisfy the canonical anti-commutation relations $\{\hat{\Phi}^{\dagger}_{n}, \hat{\Phi}_{m} \}=\delta_{nm}\hat{\mathds{1}}$, and
$\{\hat{\Phi}_{n}, \hat{\Phi}_{m} \}=0$, as for the lattice Schwinger model in Eq.~\eqref{eq: Schwinger lattice}. The first term represents the energy of the electric field $\hat{E}_{n}$. Notice the form $(\hat{E}_{n}-\hat{E}_{n-1})^{2}$ coming from the spatial derivative in Eq.~\eqref{1} in contrast to the standard $\hat{E}_n^2$ appearing in the lattice Schwinger model. The second term is the coupling between the vector potential $\hat{A}_{n}^{1}$ and the single-component fermion fields $\hat{\Phi}_{n}$. Finally, we have the staggered mass term for the fermions. The lattice sites take values $n=1,\ldots,N$ with $N$ the total number of sites.

In the Hamiltonian~\eqref{eq: quantum link H} we again consider the gauge field operators $\hat{L}_n$ and $\hat{\theta}_n$ given by Eq.~\eqref{eq: Quantum Model}. Here, we treat the electric field and vector potential as spin-1 systems with discrete spectra. For our quench procedures discussed in the following we have chosen bounded link variables $L_n = 0, \pm1$ to define a quantum link model~\cite{chandrasekharan1997quantum}. Our following discussion can be generalized to an arbitrary number of levels on the links. 



\begin{figure*}[t!]
\includegraphics[width=.9 \linewidth]{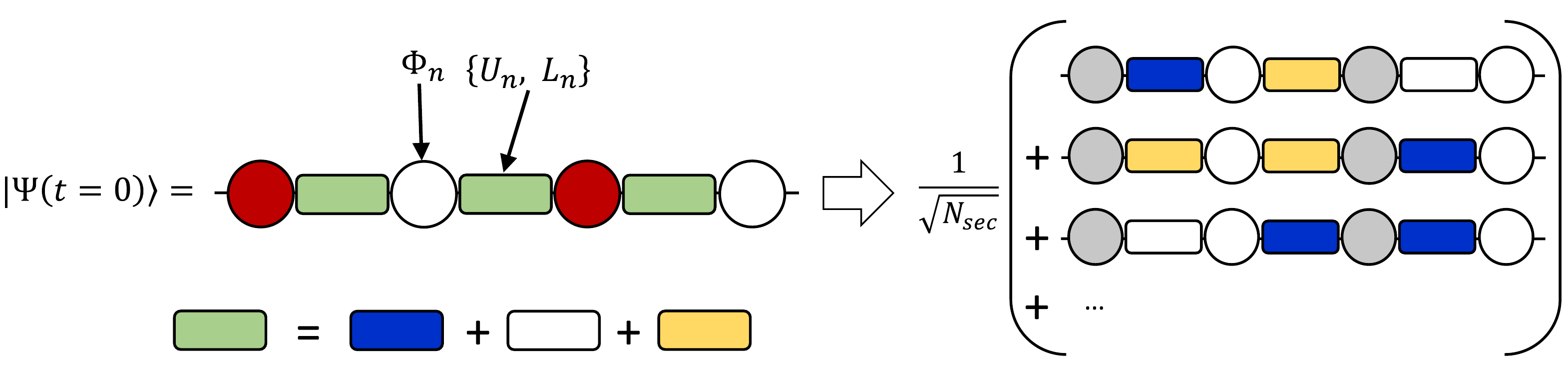}
\caption{Schematic picture of the initial state and its decomposition into $N_{sec}$ superselection sectors of charges. On the left the fermions (related to spinless operators  $\hat{\Phi}_{n}$) occupy the odd sites, with the even sites being empty. The link variables $U_{n}$ and $L_{n}$ are defined on the links between the lattice sites $n$ and $n+1$ and are related to the vector potential $\hat{A}_{n}^{1}$ and the electric field $\hat{E}_{n}$, correspondingly. The gauge fields are in an equal weight superposition of the eigenstates of the operator $\hat{L}_{n}$ with eigenvalues $L_{n}=0,\pm 1$. On the right the initial state is decomposed into $N_{sec}$ superselection sectors of charges, with the fermions maintaining their initial occupation, in the basis of the new fermionic operators $\hat{\tilde{\Phi}}_{n}$. }
\label{fig: state}
\end{figure*}

In terms of these link operators the Hamiltonian becomes
\begin{equation}\label{13}
    \begin{aligned}
    \hat{H} = \;&J_3\sum_{n}\big(\hat{L}_{n}-\hat{L}_{n-1}\big)^{2}\\
    &-i J_2 \sum_{n}\Big(\hat{\Phi}_{n}^{\dagger}e^{i\hat{\theta}_{n}}\hat{\Phi}_{n+1}-h.c.\Big)\\
    &+ J_1\sum_{n}(-1)^{n}\hat{\Phi}^{\dagger}_{n}\hat{\Phi}_{n},
    \end{aligned}
\end{equation}
where $J_3=\frac{g^{2}B}{a}, J_2 = \frac{1}{2a}$, and $J_1 = m$. The above Hamiltonian is similar to that of the lattice Schwinger model using the Kogut-Susskind formulation~\cite{kogut1975j}, with a change in the term related to the energy stored by the electric field. 

The Hamiltonian of Eq.~\eqref{13} is gauge invariant and commutes with the generators of $U(1)$ gauge symmetry, $\hat{G}_{n}$: $[\hat{G}_{n},\hat{H}]=0$. The local generators $\hat{G}_{n}$ are given by~\cite{Hamer_1997}
\begin{equation}\label{5}
   \hat{G}_{n}=\hat{L}_{n}-\hat{L}_{n-1}-\hat{\Phi}^{\dagger}_{n}\hat{\Phi}_{n}+\frac{1}{2}\left(1-(-1)^{n}\right)\hat{\mathds{1}}.
\end{equation}
These local conservation laws allow us to split the Hilbert space into superselection sectors labelled by the eigenvalues $\{q_{a}\}$, i.e.,
\begin{equation}
  \hat{G}_{n}|\Psi_{\{q_{a}\}}\rangle = q_{n}|\Psi_{\{q_{a}\}}\rangle.
\end{equation}

The standard Gauss' Law requires these local charges to be zero $\hat{G}_{n}|physical \rangle =0$, which corresponds to the subspace with gauge invariant states~\cite{kogut1975j}. However, here we consider \emph{unconstrained gauge theories}~\cite{prosko2017}, and focus on states $|\Psi_{\{q_{a}\}}\rangle$ corresponding to a distribution of local vacuum charges $\{ q_{a} \}$, the charge superselection sector of the model. Here, we note that we have promoted the gauge field operators $\hat{L}_{n}$ to physical degrees of freedom. Therefore, the Hamiltonian is symmetric with respect to $\hat{G}_n$, whereas the Hilbert space of interest is not. The existence of gauge invariance for the Hamiltonian is crucially related to the emergence of disorder in the system, as will be discussed below.

In order to reveal the free-fermion solubility of the model, our goal is to eliminate the gauge field from the Hamiltonian. We achieve this by a redefinition of the fermionic fields~\cite{Sala2018},
\begin{equation}
    \begin{split}
    \hat{\tilde{\Phi}}_{n}=e^{i\sum_{m=1}^{n-1}\hat{\theta}_{m}}\hat{\Phi}_{n},
    \end{split}
\end{equation}
which satisfy the same anti-commutation relations, and by substituting $\hat{L}_{n}-\hat{L}_{n-1}=\hat{G}_{n}+\hat{\tilde{\Phi}}^{\dagger}_{n}\hat{\tilde{\Phi}}_{n}-\frac{1}{2}\big(1-(-1)^{n}\big)\hat{\mathds{1}}$ in Eq.~\eqref{5}. The Hamiltonian then becomes
\begin{equation}\label{eq: H final}
    \begin{aligned}
        \hat{H}&=J_{3}\sum_{n}\Big(\big[(-1)^{n} -1\big]\hat{\tilde{\Phi}}_{n}^{\dagger}\hat{\tilde{\Phi}}_{n}+[2\hat{G}_{n}+1]\hat{\tilde{\Phi}}_{n}^{\dagger}\hat{\tilde{\Phi}}_{n}\Big)\\
       &-iJ_{2}\sum_{n}\Big(\hat{\tilde{\Phi}}_{n}^{\dagger}\hat{\tilde{\Phi}}_{n+1}-h.c.\Big)+J_{1}\sum_{n}(-1)^{n}\hat{\tilde{\Phi}}_{n}^{\dagger}\hat{\tilde{\Phi}}_{n}.
    \end{aligned}
\end{equation}
Note, the above procedure of eliminating the gauge fields, is similar to that used in reference~\cite{Hamer_1997}, where they proceeded with a Jordan-Wigner transformation, and eliminated the gauge fields by using the Gauss law, as well as a residual gauge transformation.

The above Hamiltonian in Eq.~\eqref{eq: H final} is a translationally invariant Hamiltonian. The first term corresponds to the emergent disorder with the effective on-site potential controlled by $\hat{G}_{n}$, which are related to the conserved charges in our model. Disorder here arises when we choose a particular sector of charges $\{ q_{a} \}$. In a given charge sector the operators $\hat{G}_{n}$ can be replaced by the their eigenvalues $q_n$, leaving a quadratic free-fermion Hamiltonian for that sector. By changing the ratio $J_{3}/J_{2}$ we change the strength of the emergent disorder. The second term is the fermion hopping term and the last term is the staggered fermion mass. In the following we set the mass ($J_{1}$) to be zero for simplicity. In each charge sector, the Hamiltonian in Eq.~\eqref{eq: H final} corresponds to a free-fermion model that is exactly solvable.

\section{Emergent disorder}\label{sec: disorder}

To study the localization behaviour in our model, which is manifestly translationally invariant, we consider a global quantum quench protocol. We start from an easily prepared initial state with an inhomogeneous fermion density and a macroscopic energy density. We then evolve with the Hamiltonian~\eqref{eq: H final} and measure the density imbalance and spreading of correlations. Following Refs.~\cite{smith2017disorder,brenes2018many} we consider the initial state of the form
\begin{equation}
    \begin{split}
        |\Psi (t=0) \rangle = |\psi  \rangle_{g} \otimes |\psi  \rangle_{f}.
    \end{split}
\end{equation}
The fermionic part $|\psi  \rangle  _{f}$ is a Slater determinant with fermions on the odd sites and the even sites are not occupied. The gauge fields, defined on the bonds, are set to be in an equal weight superposition of the eigenstates of $\hat{L}_{n}$ with eigenvalues $L_{n}=0,\pm 1$. We can write
\begin{equation}\label{initialst}
    \begin{split}
        |\Psi (t=0) \rangle= \big[\otimes_{n}|\tilde{L}_{n}  \rangle \big] \otimes |1010\cdots \rangle_{f},
    \end{split}
\end{equation}
with
\begin{equation}\label{469}
    \begin{split}
       |\tilde{L}_{n}  \rangle  =\frac{1}{\sqrt{3}}\Big(|-1  \rangle  _{n}+|0  \rangle  _{n}+|1  \rangle  _{n}\Big),
    \end{split}
\end{equation}
for $n=1,...,N-1$. The initial state can be rewritten as
\begin{equation}
    \begin{split}
      |\Psi (t=0) \rangle =\frac{1}{\sqrt{N_{sec}}}\sum_{\{q_{a}\}}| \{q_a\} \rangle \otimes |1010\cdots  \rangle_{\tilde{f}},
    \end{split}
\end{equation}
where $N_{sec}$ is the total number of superselection sectors of charges, and  the fermions are in the basis of $\hat{\tilde{\Phi}}$ operators. The local charges $q_{n}$ of each sector are given by Eq.~\eqref{5} and depend on the distribution of the fermions and the gauge fields. Specifically, each superselection sector corresponds to a particular distribution of the gauge fields with local eigenvalues $L_{n}=0,\pm 1$.  Schematically, the decomposition of the initial state is shown in Fig.~\ref{fig: state}.
The time evolution of the state is then given by
\begin{equation}
    \begin{aligned}
  |\Psi(t)\rangle = \frac{1}{\sqrt{N_{sec}}}\sum_{\{q_{a}\}}e^{-it\hat{H}_{\{q_{a}\}}}|\{q_{a}\} \rangle \otimes |\psi\rangle_{\tilde{f}},
    \end{aligned}
\end{equation}
where in the Hamiltonians $\hat{H}_{\{q_{a}\}}$ the generators $\hat{G}_n$ are replaced by the local charges $q_n$ that correspond to a particular charge sector. The Hamiltonian $\hat{H}_{\{q_a\}}$ contains a disorder potential which is controlled by the local charges of the specific charge sector ${\{q_a\}}$. We consider the dynamics of the density imbalance between odd and even sites,
\begin{equation}\label{241}
\begin{split}
\Delta \rho (t)=\frac{1}{\tilde{N}}\sum_{j=1}^{N-1} |\langle \Psi(t) |\big(\hat{n}_{j}-\hat{n}_{j+1}\big)|\Psi(t) \rangle |,
\end{split}
\end{equation}
and the connected density correlator
\begin{equation}\label{eq: density correlator}
\begin{aligned}
\langle &\Psi (t)|\hat{n}_{j}\hat{n}_{k}|\Psi (t)\rangle_{c}\\
&= \langle \Psi (t)|\hat{n}_{j}\hat{n}_{k}|\Psi (t)\rangle
-\langle \Psi (t)|\hat{n}_{j}|\Psi (t)\rangle \langle \Psi (t)|\hat{n}_{k}|\Psi (t)\rangle,
\end{aligned}
\end{equation}
where $\tilde{N} = N-1$ with $N$ the number of sites and $\hat{n}_j = \hat{\tilde{\Phi}}_j^\dag \hat{\tilde{\Phi}}_j$. These quantities allow us to diagnose the localization behaviour in our model signalled by the persistence of the initial density balance~\cite{Hauschild2016,Choi2016,Schreiber2015} and the lack of correlation spreading after the quench~\cite{Smith2018}.

\begin{figure}[t]
\centering
\includegraphics[width=\columnwidth]{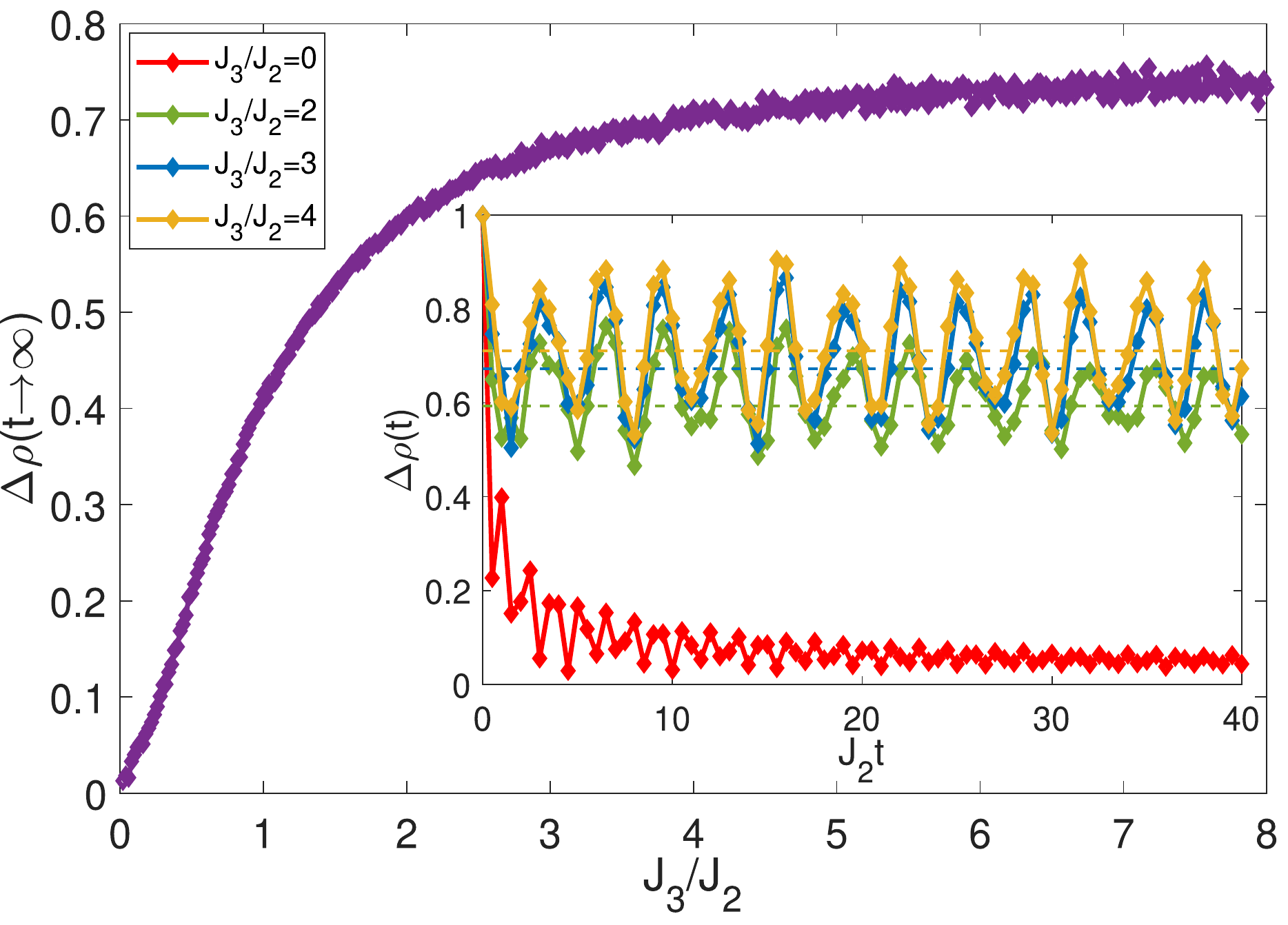}
\caption{The asymptotic behaviour of the average density imbalance (computed at $J_{2}t=10^{6})$ as a function of $J_{3}/J_{2}$, with $J_{1}=0$. The lattice contains 100 lattice sites and we averaged over 200 superselection sectors. We also plot (inset) how the average density imbalance changes with time, for four values of $J_{3}/J_{2}$. As $J_{3}/J_{2}$ increases, memory of the initial state is more apparent. Dash lines indicate the asymptotic behaviour. 
}\label{fig: density imbalance 1}
\end{figure}

\begin{figure}[b!] 
\includegraphics[width=\columnwidth]{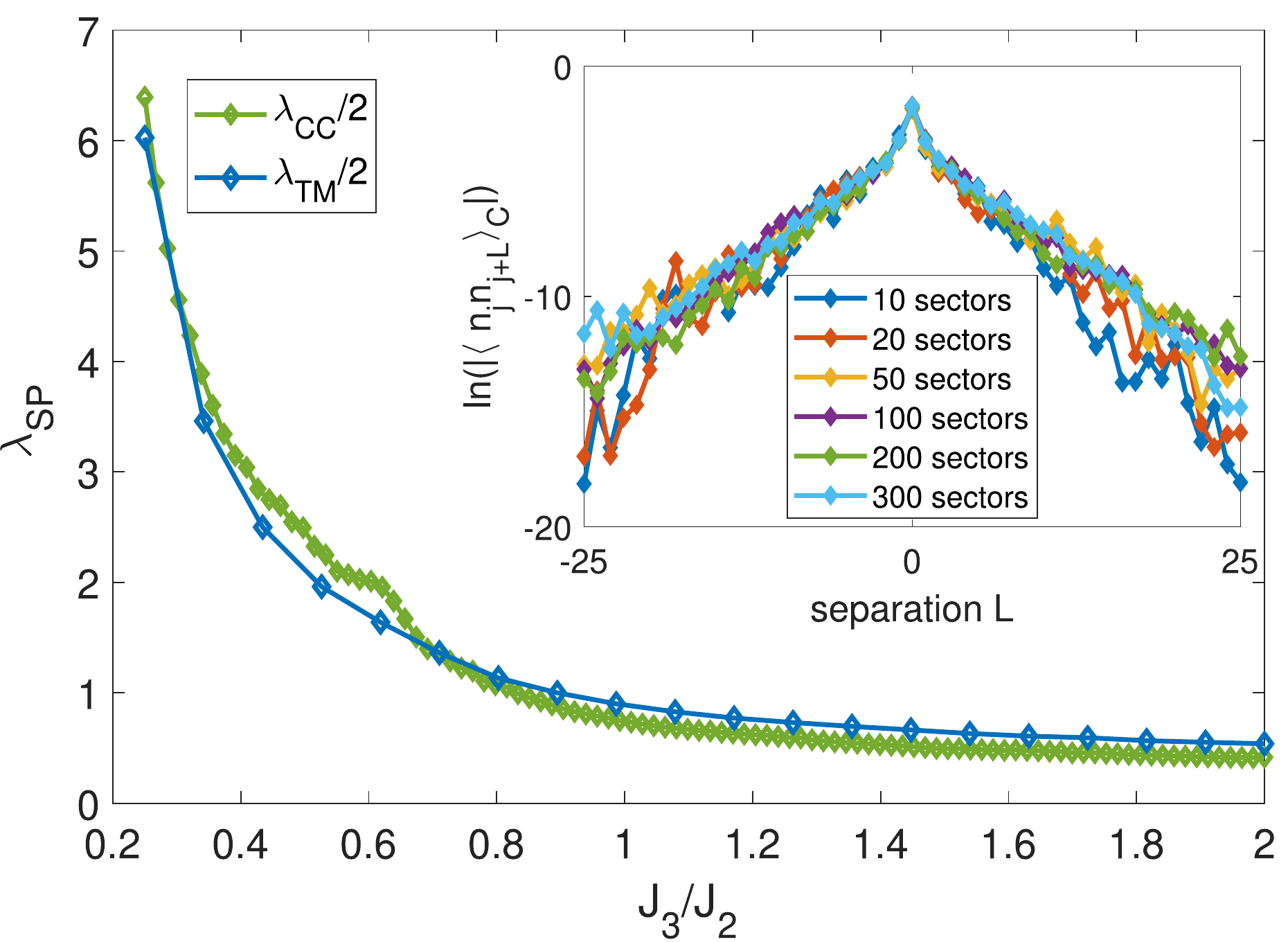}
\centering
\caption{The localization length $\lambda_{CC}$ for different disorder strengths $J_{3}/J_{2}$, extracted from the density correlator at $t=100/J_{2}$.  The lattice contains 200 sites and the number of superselection sectors of charges is 200. We also plot the localization length obtained from the Transfer Matrix approach $\lambda_{TM}$ for a lattice of $10^{5}$ sites. We plot (inset) the long time limit ($t=10^{9}/J_{2}$) of the absolute value of the connected density correlator on a log scale for different numbers of superselection sectors of charges and for disorder strength $J_{3}/J_{2}=1$. We show convergence as approaching 300 superselection sectors of charges.}\label{fig: loc length}
\end{figure}

In order to compute these observables we note that the density imbalance can be written as
\begin{equation}
    \begin{aligned}
        \Delta &\rho (t)\\ &=\frac{1}{\tilde{N}N_{sec}}\sum_{j}\sum_{\{q_{a}\}} |\langle \psi (t) |_{\tilde{f}} \big(\hat{n}_{j}-\hat{n}_{j+1}\big) |\psi (t) \rangle_{\tilde{f}} |,
    \end{aligned}
\end{equation}
where $|\psi(t)\rangle_{\tilde{f}} = e^{-it\hat{H}_{\{q_a\}}}|\psi\rangle_{\tilde{f}}$ is the fermion state evolved under the Hamiltonian in a given charge sector.  Here, we have a disorder average over all possible sectors of charges even though our state is translationally invariant. The density correlator can similarly be written as an average of free-fermion correlators. We calculated the average density imbalance numerically using free-fermion methods, as summarized in Ref.~\cite{Smith2018}. We average over a limited number of superselection sectors and find that the system is self-averaging and a small fraction of the total number of sectors is sufficient for converged results, see Fig.~\ref{fig: loc length}(inset).\\

In Fig.~\ref{fig: density imbalance 1} we show the result of the density imbalance. Initially $\Delta\rho(t=0)=1$ but at long time we expect $\Delta\rho(t\rightarrow \infty)=0$ for an ergodic system, which corresponds to a uniform fermion density. For our model we find that $\Delta\rho(t\rightarrow \infty)\neq0$ for all non-zero values of $J_3/J_2$. This indicates that the system has long-time memory of the initial state and the remaining imbalance increases monotonically with $J_3/J_2$, as shown in Fig.~\ref{fig: density imbalance 1}. We compare this with the case with no effective disorder $J_3=0$, where the density imbalance decays to zero at long-times (red curve) in the thermodynamic limit. We note that the observed power-law decay is due to the integrability of the model.

\begin{figure*}[!] 
\includegraphics[width=1 \linewidth]{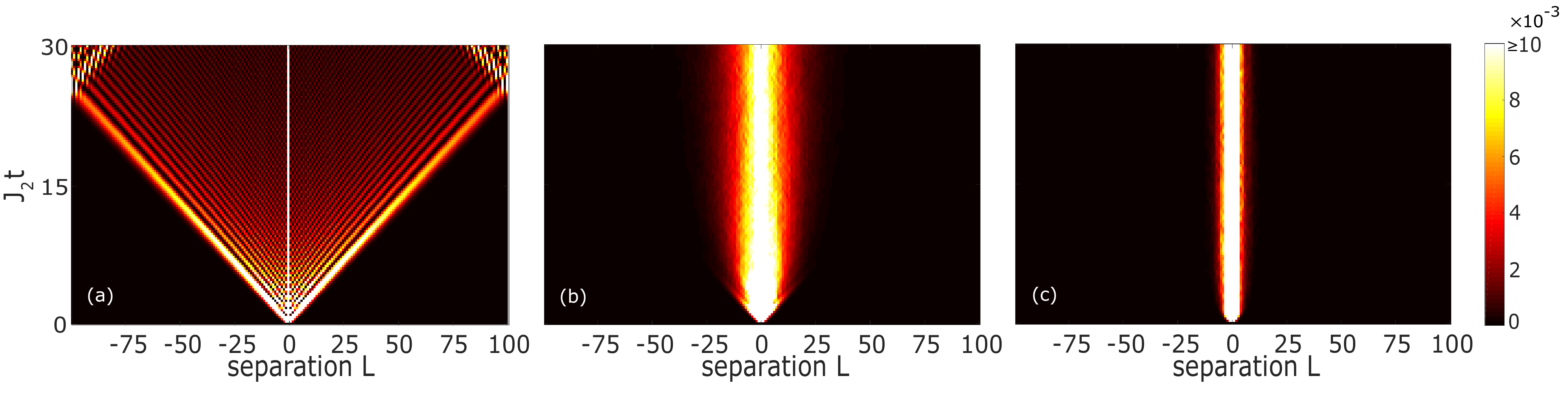}
\centering
        \caption{The absolute value of the connected density correlator $|\langle \Psi (t)|\hat{n}_{j}\hat{n}_{j+L}|\Psi (t)\rangle_{c}|$ is shown for a lattice of 200 sites. (a) $J_{3}/J_{2}=0$, leading to a linear light-cone. Two with non-zero effective disorder strengths are shown in (b) $J_{3}/J_{2}=0.3$ and (c) $J_{3}/J_{2}=1$. We average over 200 charge superselection sectors.}
        \label{fig:three graphs}
\end{figure*}
\begin{figure}[h!] 
\includegraphics[width=\columnwidth]{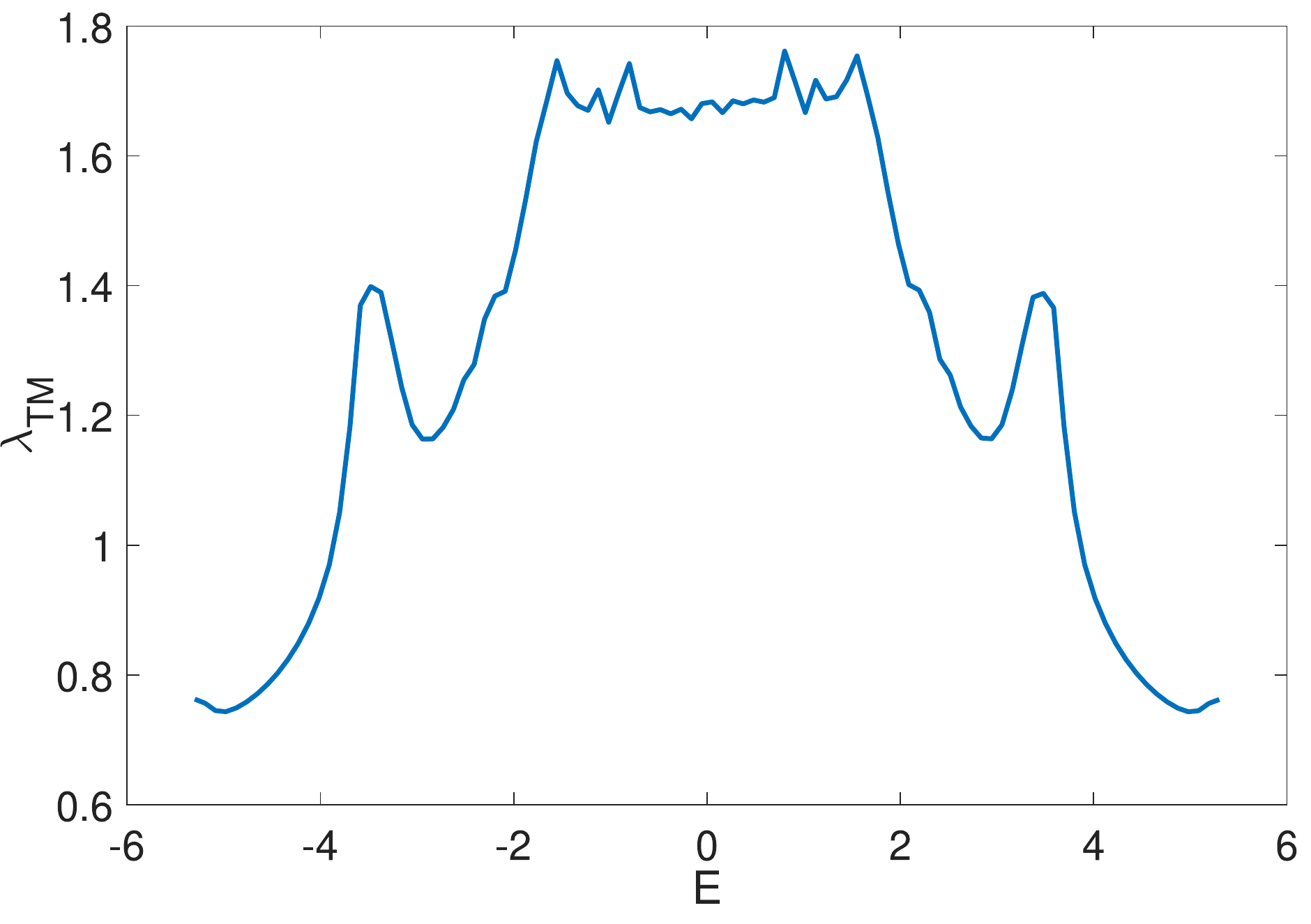}
\centering
\caption{The energy-resolved localization length obtained from the Transfer Matrix approach $\lambda_{TM}$, for disorder strength $J_{3}/J_{2}=1$, for a specific sector of charges. The lattice contains $10^{5}$ sites.}\label{fig: loc length2}
\end{figure}

Next we consider the spreading of the density correlations, defined in Eq.~\eqref{eq: density correlator} and shown in Fig.~\ref{fig:three graphs}. We show the free case $J_3 = 0$ in Fig.~\ref{fig:three graphs}(a). The density correlations can be seen to spread ballistically until they reach the edge of the system. In comparison, for non-zero $J_3/J_2$ we find that the spreading of correlations is quickly halted and the correlations have a finite extent at long times, as shown in Figs.~\ref{fig:three graphs}(b) and ~\ref{fig:three graphs}(c). This lack of spreading is yet another piece of evidence for the localization behaviour in the model.

At long times the spatial profile of the density correlations decay exponentially, as shown in the inset of Fig.~\ref{fig: loc length}. This exponential decay is related to the localization of the single particle wavefunctions. In Ref.~\cite{smith2017disorder} it was found that this profile is given by $\exp (-L/\lambda_{CC})=\exp (-L/2\lambda_{SP})$, where $\lambda_{SP}$ is the single-particle localization length. The localization length of a single particle eigenstate $\lambda(E)$ is the length scale for the exponential envelope $\sim e^{-|j-i|/\lambda(E)}$ for that state, and $\lambda_{SP}$ is the maximum such localization length averaged over disorder realisations. The localization length extracted from the density correlator is shown in Fig.~\ref{fig: loc length}, which decreases with increasing effective disorder strength $J_3/J_2$. We also plot in Fig.~\ref{fig: loc length} the localization length obtained using the Transfer Matrix approach which is in good agreement with the one extracted from the correlator. Finally, in Fig.~\ref{fig: loc length2} we plot the energy-resolved localization length for disorder strength $J_{3}/J_{2}=1$ obtained from the Transfer Matrix approach for a system of $10^{5}$ lattice sites~\cite{smith2019localization}. It shows that despite correlations in the disorder potential, the entire spectrum of eigenstates is localized.

\section{Our Model: Wilson Fermions}\label{sec: wilson}

Having shown the localization behaviour in a simple LGT using staggered fermions we now consider an alternative discretization in terms of Wilson fermions~\cite{wilson1977new}, namely
\begin{equation}\label{eq: Wilson}
    \begin{aligned}
\hat{H}_{W}=&\frac{B}{a} \sum_{n} \big(\hat{E}_{n}-\hat{E}_{n-1}\big)^{2}\\
&+a\sum_{n}\Big(\hat{\Psi}_{n}^{\dagger}\gamma^{0}\Big[m+\frac{r}{a}\Big]\hat{\Psi}_{n}\Big)\\
&-\frac{1}{2}\sum_{n}\Big(\hat{\Psi}_{n}^{\dagger}\gamma^{0}\Big[i\gamma^{1}+r\Big]\hat{U}_{n}\hat{\Psi}_{n+1}+h.c.\Big),
    \end{aligned}
\end{equation}	
where $\hat{\Psi}_n$ are two-component spinors. The above Hamiltonian gives the continuum limit of Eq.~\eqref{1} as $a\rightarrow 0$ and $r$ is the Wilson parameter which can take any value in the interval r $\in (0,1]$ without changing the continuum limit of the theory. $\hat{U}_{n}$ are the $U(1)$ parallel transporters defined as $\hat{U}_{n}:=e^{-iag\hat{A}_{n}^{1}}$. Furthermore, Eq.~\eqref{eq: Wilson} is gauge invariant, with the generators related to the gauge transformations $\hat{G}_{n}$, being defined as as~\cite{zache2018quantum}
 \begin{equation}\label{2}
    \begin{gathered}
    \hat{G}_{n}=\hat{E}_{n}-\hat{E}_{n-1}-g\big(\hat{\Psi}_{n}^{\dagger} \hat{\Psi}_{n}-1\big).
    \end{gathered}
\end{equation}  
With the aim to rewrite the Hamiltonian in terms of component fields we choose a particular representation of the gamma matrices
  \begin{equation}
    \begin{gathered}
    \gamma^{0}=\begin{pmatrix}0 & 1\\ 1 & 0 \end{pmatrix} \qquad 
    \gamma^{1}=\begin{pmatrix}i & 0\\ 0 & -i \end{pmatrix}.
        \end{gathered}
        \end{equation}
        If we write the upper and lower component of the Dirac spinor $\hat{\Psi}_{n}$ as $\hat{\Psi}_{n,1}$ and $\hat{\Psi}_{n,2}$ respectively, the Hamiltonian of Eq.~\eqref{eq: Wilson} can be written on the lattice as
\begin{equation}\label{51}
    \begin{aligned}
\hat{H}= &\frac{B}{a}\sum_{n} \big(\hat{E}_{n}-\hat{E}_{n-1}\big)^{2}\\
&+\sum_{n}\Big(m+\frac{1}{a}\Big)\big(\hat{\Psi}_{n,1}^{\dagger}\hat{\Psi}_{n,2}+\hat{\Psi}_{n,2}^{\dagger}\hat{\Psi}_{n,1}\big)\\
&+\frac{1}{a}\sum_{n}\Big(\hat{\Psi}_{n,1}^{\dagger}\hat{U}_{n}\hat{\Psi}_{n+1,2}+h.c. \Big),
    \end{aligned}
\end{equation}
where we redefined the fermionic fields as $\hat{\Psi}_{n,i}\rightarrow (-1)^{n}\sqrt{a}\hat{\Psi}_{n,i}$ and set the Wilson parameter to $r=1$. The result of the above field redefinition, is that the fermionic fields now are dimensionless. We followed the same approach as Ref.~\cite{zache2018quantum}, writing the Hamiltonian in a convenient form. These operators satisfy the following commutation and anti-commutation relations
\begin{equation}
    \begin{aligned}
\{\hat{\Psi}_{n,\alpha},\hat{\Psi}^{\dagger}_{m,\beta} \}=\delta_{\alpha\beta}\delta_{nm}\hat{\mathds{1}}\\
[\hat{E}_{n},\hat{U}_{m}]=g\delta_{nm}\hat{U}_{m}.
    \end{aligned}
\end{equation}	
Let us now follow the similar steps as before for the Staggered fermions studied in Sec.~\ref{sec: continuum}. We introduce the operators $\hat{\theta}_{n}$ and $\hat{L}_{n}$ as defined in Eq. \eqref{eq: Quantum Model}. Equation~\eqref{51} then becomes
 \begin{equation}
    \begin{aligned}
\hat{H}= &\frac{Bg^{2}}{a}\sum_{n} \big(\hat{L}_{n}-\hat{L}_{n-1}\big)^{2}\\
&+\sum_{n}\Big(m+\frac{1}{a}\Big)\big(\hat{\Psi}_{n,1}^{\dagger}\hat{\Psi}_{n,2}+\hat{\Psi}_{n,2}^{\dagger}\hat{\Psi}_{n,1}\big)\\
&+\frac{1}{a}\sum_{n}\Big(\hat{\Psi}_{n,1}^{\dagger}\hat{U}_{n}\hat{\Psi}_{n+1,2}+h.c. \Big),
    \end{aligned}
\end{equation}
 with $\hat{U}_{n}=e^{i\hat{\theta}_{n}}$, and the generators $\hat{G}_{n}$ take the form:
 \begin{equation}\label{eq: generators2}
    \begin{gathered}
    \hat{G}_{n}=g\hat{L}_{n}-g\hat{L}_{n-1}-g\big(\hat{\Psi}_{n}^{\dagger}\hat{\Psi}_{n}-1\big).
    \end{gathered}
\end{equation} 
The next step is to redefine the fermionic fields
\begin{equation}
\begin{gathered}
\hat{\tilde{\Psi}}_{n,j}=e^{i\sum_{l=1}^{n-1}\hat{\theta}_l}\hat{\Psi}_{n,j}\\
\end{gathered}
\end{equation} 
for $j=1,2$ and subtitute Eq.~\eqref{eq: generators2} into the Hamiltonian to get
\begin{equation}
    \begin{gathered}
    \hat{H}_{W}=\frac{B}{a}\sum_{n}\Big(\hat{G}_{n}-g\big[1-\hat{\tilde{\Psi}}^{\dagger}_{n,1}\hat{\tilde{\Psi}}_{n,1}-\hat{\tilde{\Psi}}^{\dagger}_{n,2}\hat{\tilde{\Psi}}_{n,2}\big]\Big)^{2}\\+\Big(m+\frac{1}{a}\Big)\sum_{n}\big(\hat{\tilde{\Psi}}_{n,1}^{\dagger}\hat{\tilde{\Psi}}_{n,2}+\hat{\tilde{\Psi}}_{n,2}^{\dagger}\hat{\tilde{\Psi}}_{n,1}\big)
    \\+\frac{1}{a}\sum_{n}\big(\hat{\tilde{\Psi}}_{n,1}^{\dagger}\hat{\tilde{\Psi}}_{n+1,2}+h.c.\big).
    \end{gathered}
\end{equation}
 In contrast to the staggered fermions, the first term contains both a disordered potential and a fermionic interaction term, namely
\begin{equation}
    \begin{split}
    \hat{H}_{int}=\frac{B}{a}\sum_{n}2g^{2}\big(\hat{\tilde{\Psi}}^{\dagger}_{n,1}\hat{\tilde{\Psi}}_{n,1}\hat{\tilde{\Psi}}^{\dagger}_{n,2}\hat{\tilde{\Psi}}_{n,2}\big).
    \end{split}
\end{equation}
Interactions lead to a model that is not free-fermion soluble and a numerical study of its properties is beyond the scope of this paper.

\section{Discussion}

In this paper we have studied a simple $U(1)$ lattice gauge theory derived from a variant of continuum QED in $(1+1)$ dimension. By changing only the dynamics of the gauge field we remove the effective long-range interactions that result in confinement and reveal emergent disorder coming from the local symmetry of the model. We focused on the staggered fermion formulation of the lattice model where the model was free-fermion soluble in analogy to the Z$_2$ LGT of Ref.~\cite{smith2017disorder} and, therefore, amenable to large scale numerical simulations proving the disorder-free localization mechanism. Surprisingly, the Wilson fermion formulation results in a model that is manifestly interacting. In this setting it is then a question whether the lattice model is many-body localized.

We have shown that in the disorder-free localization mechanism the emergent disorder potential depends on the distribution of the fermion and gauge field configuration of the initial state Eq.~\eqref{5}. For our choice of quench set-up the resulting disorder was only weakly correlated leading to full localization shown in Fig.~\ref{fig: loc length2}. An interesting topic for future research would be a systematic study of whether special choices of initial states can give rise to correlated disorder, which in principle could result in mobility edges or delocalization even in one dimension. 

Another extension of this work would be to consider different gauge fields, in particular non-abelian gauge theories, as well as studying higher-dimensional models. In these settings, an open question is whether continuum gauge theories that have soluble discretizations exist and whether the disorder-free mechanism for localization extends more generally. 

Finally, our simple modification of $(1+1)$-dimensional quantum electrodynamics allowed us to find a basic free fermion solution, which in turn could be of potential interest for the study of dynamical phenomena  in the gauge invariant sector of the model like pair-production protocols.  

\begin{acknowledgements}
We thank Dmitry Kovrizhin and Roderich Moessner for previous collaborations related to this work. We are very grateful to Pablo Sala for reading drafts of the paper and helpful comments.
A.S. was supported by the European Research Council (ERC) under the European Union's Horizon 2020 research and innovation programme (grant agreement No. 771537).
\end{acknowledgements}


\begin{thebibliography}{42}%
	\makeatletter
	\providecommand \@ifxundefined [1]{%
		\@ifx{#1\undefined}
	}%
	\providecommand \@ifnum [1]{%
		\ifnum #1\expandafter \@firstoftwo
		\else \expandafter \@secondoftwo
		\fi
	}%
	\providecommand \@ifx [1]{%
		\ifx #1\expandafter \@firstoftwo
		\else \expandafter \@secondoftwo
		\fi
	}%
	\providecommand \natexlab [1]{#1}%
	\providecommand \enquote  [1]{``#1''}%
	\providecommand \bibnamefont  [1]{#1}%
	\providecommand \bibfnamefont [1]{#1}%
	\providecommand \citenamefont [1]{#1}%
	\providecommand \href@noop [0]{\@secondoftwo}%
	\providecommand \href [0]{\begingroup \@sanitize@url \@href}%
	\providecommand \@href[1]{\@@startlink{#1}\@@href}%
	\providecommand \@@href[1]{\endgroup#1\@@endlink}%
	\providecommand \@sanitize@url [0]{\catcode `\\12\catcode `\$12\catcode
		`\&12\catcode `\#12\catcode `\^12\catcode `\_12\catcode `\%12\relax}%
	\providecommand \@@startlink[1]{}%
	\providecommand \@@endlink[0]{}%
	\providecommand \url  [0]{\begingroup\@sanitize@url \@url }%
	\providecommand \@url [1]{\endgroup\@href {#1}{\urlprefix }}%
	\providecommand \urlprefix  [0]{URL }%
	\providecommand \Eprint [0]{\href }%
	\providecommand \doibase [0]{http://dx.doi.org/}%
	\providecommand \selectlanguage [0]{\@gobble}%
	\providecommand \bibinfo  [0]{\@secondoftwo}%
	\providecommand \bibfield  [0]{\@secondoftwo}%
	\providecommand \translation [1]{[#1]}%
	\providecommand \BibitemOpen [0]{}%
	\providecommand \bibitemStop [0]{}%
	\providecommand \bibitemNoStop [0]{.\EOS\space}%
	\providecommand \EOS [0]{\spacefactor3000\relax}%
	\providecommand \BibitemShut  [1]{\csname bibitem#1\endcsname}%
	\let\auto@bib@innerbib\@empty
	\bibitem [{\citenamefont {Anderson}(1958)}]{anderson1958absence}%
	\BibitemOpen
	\bibfield  {author} {\bibinfo {author} {\bibfnamefont {P.~W.}\ \bibnamefont
			{Anderson}},\ }\href {\doibase 10.1103/PhysRev.109.1492} {\bibfield
		{journal} {\bibinfo  {journal} {Phys. Rev.}\ }\textbf {\bibinfo {volume}
			{109}},\ \bibinfo {pages} {1492} (\bibinfo {year} {1958})}\BibitemShut
	{NoStop}%
	\bibitem [{\citenamefont {Scardicchio}\ and\ \citenamefont
		{Thiery}(2017)}]{scardicchio2017perturbation}%
	\BibitemOpen
	\bibfield  {author} {\bibinfo {author} {\bibfnamefont {A.}~\bibnamefont
			{Scardicchio}}\ and\ \bibinfo {author} {\bibfnamefont {T.}~\bibnamefont
			{Thiery}},\ }\href@noop {} {\bibfield  {journal} {\bibinfo  {journal} {arXiv
				preprint arXiv:1710.01234}\ } (\bibinfo {year} {2017})}\BibitemShut {NoStop}%
	\bibitem [{\citenamefont {Basko}\ \emph {et~al.}(2006)\citenamefont {Basko},
		\citenamefont {Aleiner},\ and\ \citenamefont {Altshuler}}]{basko2006metal}%
	\BibitemOpen
	\bibfield  {author} {\bibinfo {author} {\bibfnamefont {D.~M.}\ \bibnamefont
			{Basko}}, \bibinfo {author} {\bibfnamefont {I.~L.}\ \bibnamefont {Aleiner}},
		\ and\ \bibinfo {author} {\bibfnamefont {B.~L.}\ \bibnamefont {Altshuler}},\
	}\href@noop {} {\bibfield  {journal} {\bibinfo  {journal} {Annals of
				physics}\ }\textbf {\bibinfo {volume} {321}},\ \bibinfo {pages} {1126}
		(\bibinfo {year} {2006})}\BibitemShut {NoStop}%
	\bibitem [{\citenamefont {{\v{Z}}nidari{\v{c}}}\ \emph
		{et~al.}(2008)\citenamefont {{\v{Z}}nidari{\v{c}}}, \citenamefont {Prosen},\
		and\ \citenamefont {Prelov{\v{s}}ek}}]{Znidaric2008}%
	\BibitemOpen
	\bibfield  {author} {\bibinfo {author} {\bibfnamefont {M.}~\bibnamefont
			{{\v{Z}}nidari{\v{c}}}}, \bibinfo {author} {\bibfnamefont {T.}~\bibnamefont
			{Prosen}}, \ and\ \bibinfo {author} {\bibfnamefont {P.}~\bibnamefont
			{Prelov{\v{s}}ek}},\ }\href {\doibase 10.1103/PhysRevB.77.064426} {\bibfield
		{journal} {\bibinfo  {journal} {Phys. Rev. B}\ }\textbf {\bibinfo {volume}
			{77}},\ \bibinfo {pages} {064426} (\bibinfo {year} {2008})}\BibitemShut
	{NoStop}%
	\bibitem [{\citenamefont {Bardarson}\ \emph {et~al.}(2012)\citenamefont
		{Bardarson}, \citenamefont {Pollmann},\ and\ \citenamefont
		{Moore}}]{Bardason2012}%
	\BibitemOpen
	\bibfield  {author} {\bibinfo {author} {\bibfnamefont {J.~H.}\ \bibnamefont
			{Bardarson}}, \bibinfo {author} {\bibfnamefont {F.}~\bibnamefont {Pollmann}},
		\ and\ \bibinfo {author} {\bibfnamefont {J.~E.}\ \bibnamefont {Moore}},\
	}\href {\doibase 10.1103/PhysRevLett.109.017202} {\bibfield  {journal}
		{\bibinfo  {journal} {Phys. Rev. Lett.}\ }\textbf {\bibinfo {volume} {109}},\
		\bibinfo {pages} {017202} (\bibinfo {year} {2012})}\BibitemShut {NoStop}%
	\bibitem [{\citenamefont {Abanin}\ and\ \citenamefont
		{Papi{\'{c}}}(2017)}]{abanin2017}%
	\BibitemOpen
	\bibfield  {author} {\bibinfo {author} {\bibfnamefont {D.~A.}\ \bibnamefont
			{Abanin}}\ and\ \bibinfo {author} {\bibfnamefont {Z.}~\bibnamefont
			{Papi{\'{c}}}},\ }\href {\doibase 10.1002/andp.201700169} {\bibfield
		{journal} {\bibinfo  {journal} {Ann. Phys.}\ }\textbf {\bibinfo {volume}
			{529}},\ \bibinfo {pages} {1700169} (\bibinfo {year} {2017})}\BibitemShut
	{NoStop}%
	\bibitem [{\citenamefont {Nandkishore}\ and\ \citenamefont
		{Huse}(2015)}]{nandkishore2015many}%
	\BibitemOpen
	\bibfield  {author} {\bibinfo {author} {\bibfnamefont {R.}~\bibnamefont
			{Nandkishore}}\ and\ \bibinfo {author} {\bibfnamefont {D.~A.}\ \bibnamefont
			{Huse}},\ }\href@noop {} {\bibfield  {journal} {\bibinfo  {journal} {Annu.
				Rev. Condens. Matter Phys.}\ }\textbf {\bibinfo {volume} {6}},\ \bibinfo
		{pages} {15} (\bibinfo {year} {2015})}\BibitemShut {NoStop}%
	\bibitem [{\citenamefont {D'Alessio}\ \emph {et~al.}(2016)\citenamefont
		{D'Alessio}, \citenamefont {Kafri}, \citenamefont {Polkovnikov},\ and\
		\citenamefont {Rigol}}]{DAlessio2016}%
	\BibitemOpen
	\bibfield  {author} {\bibinfo {author} {\bibfnamefont {L.}~\bibnamefont
			{D'Alessio}}, \bibinfo {author} {\bibfnamefont {Y.}~\bibnamefont {Kafri}},
		\bibinfo {author} {\bibfnamefont {A.}~\bibnamefont {Polkovnikov}}, \ and\
		\bibinfo {author} {\bibfnamefont {M.}~\bibnamefont {Rigol}},\ }\href
	{\doibase 10.1080/00018732.2016.1198134} {\bibfield  {journal} {\bibinfo
			{journal} {Adv. Phys.}\ }\textbf {\bibinfo {volume} {65}},\ \bibinfo {pages}
		{239} (\bibinfo {year} {2016})}\BibitemShut {NoStop}%
	\bibitem [{\citenamefont {Deutsch}(1991)}]{Deutsch1991}%
	\BibitemOpen
	\bibfield  {author} {\bibinfo {author} {\bibfnamefont {J.~M.}\ \bibnamefont
			{Deutsch}},\ }\href {\doibase 10.1103/PhysRevA.43.2046} {\bibfield  {journal}
		{\bibinfo  {journal} {Phys. Rev. A}\ }\textbf {\bibinfo {volume} {43}},\
		\bibinfo {pages} {2046} (\bibinfo {year} {1991})}\BibitemShut {NoStop}%
	\bibitem [{\citenamefont {Srednicki}(1994)}]{Srednicki1994}%
	\BibitemOpen
	\bibfield  {author} {\bibinfo {author} {\bibfnamefont {M.}~\bibnamefont
			{Srednicki}},\ }\href {\doibase 10.1103/PhysRevE.50.888} {\bibfield
		{journal} {\bibinfo  {journal} {Phys. Rev. E}\ }\textbf {\bibinfo {volume}
			{50}},\ \bibinfo {pages} {888} (\bibinfo {year} {1994})}\BibitemShut
	{NoStop}%
	\bibitem [{\citenamefont {Smith}\ \emph {et~al.}(2017)\citenamefont {Smith},
		\citenamefont {Knolle}, \citenamefont {Kovrizhin},\ and\ \citenamefont
		{Moessner}}]{smith2017disorder}%
	\BibitemOpen
	\bibfield  {author} {\bibinfo {author} {\bibfnamefont {A.}~\bibnamefont
			{Smith}}, \bibinfo {author} {\bibfnamefont {J.}~\bibnamefont {Knolle}},
		\bibinfo {author} {\bibfnamefont {D.~L.}\ \bibnamefont {Kovrizhin}}, \ and\
		\bibinfo {author} {\bibfnamefont {R.}~\bibnamefont {Moessner}},\ }\href
	{\doibase 10.1103/PhysRevLett.118.266601} {\bibfield  {journal} {\bibinfo
			{journal} {Phys. Rev. Lett.}\ }\textbf {\bibinfo {volume} {118}},\ \bibinfo
		{pages} {266601} (\bibinfo {year} {2017})}\BibitemShut {NoStop}%
	\bibitem [{\citenamefont {Prosko}\ \emph {et~al.}(2017)\citenamefont {Prosko},
		\citenamefont {Lee},\ and\ \citenamefont {Maciejko}}]{prosko2017}%
	\BibitemOpen
	\bibfield  {author} {\bibinfo {author} {\bibfnamefont {C.}~\bibnamefont
			{Prosko}}, \bibinfo {author} {\bibfnamefont {S.-P.}\ \bibnamefont {Lee}}, \
		and\ \bibinfo {author} {\bibfnamefont {J.}~\bibnamefont {Maciejko}},\ }\href
	{\doibase 10.1103/PhysRevB.96.205104} {\bibfield  {journal} {\bibinfo
			{journal} {Phys. Rev. B}\ }\textbf {\bibinfo {volume} {96}},\ \bibinfo
		{pages} {205104} (\bibinfo {year} {2017})}\BibitemShut {NoStop}%
	\bibitem [{\citenamefont {Brenes}\ \emph {et~al.}(2018)\citenamefont {Brenes},
		\citenamefont {Dalmonte}, \citenamefont {Heyl},\ and\ \citenamefont
		{Scardicchio}}]{brenes2018many}%
	\BibitemOpen
	\bibfield  {author} {\bibinfo {author} {\bibfnamefont {M.}~\bibnamefont
			{Brenes}}, \bibinfo {author} {\bibfnamefont {M.}~\bibnamefont {Dalmonte}},
		\bibinfo {author} {\bibfnamefont {M.}~\bibnamefont {Heyl}}, \ and\ \bibinfo
		{author} {\bibfnamefont {A.}~\bibnamefont {Scardicchio}},\ }\href {\doibase
		10.1103/PhysRevLett.120.030601} {\bibfield  {journal} {\bibinfo  {journal}
			{Phys. Rev. Lett.}\ }\textbf {\bibinfo {volume} {120}},\ \bibinfo {pages}
		{030601} (\bibinfo {year} {2018})}\BibitemShut {NoStop}%
	\bibitem [{\citenamefont {Karpov}\ \emph {et~al.}(2020)\citenamefont {Karpov},
		\citenamefont {Verdel}, \citenamefont {Huang}, \citenamefont {Schmitt},\ and\
		\citenamefont {Heyl}}]{Karpov2020}%
	\BibitemOpen
	\bibfield  {author} {\bibinfo {author} {\bibfnamefont {P.}~\bibnamefont
			{Karpov}}, \bibinfo {author} {\bibfnamefont {R.}~\bibnamefont {Verdel}},
		\bibinfo {author} {\bibfnamefont {Y.-P.}\ \bibnamefont {Huang}}, \bibinfo
		{author} {\bibfnamefont {M.}~\bibnamefont {Schmitt}}, \ and\ \bibinfo
		{author} {\bibfnamefont {M.}~\bibnamefont {Heyl}},\ }\href@noop {} {\emph
		{\bibinfo {title} {{Disorder-free localization in an interacting
					two-dimensional lattice gauge theory}}}},\ \bibinfo {type} {Tech. Rep.}\
	(\bibinfo {year} {2020})\ \Eprint {http://arxiv.org/abs/2003.04901v1}
	{arXiv:2003.04901v1} \BibitemShut {NoStop}%
	\bibitem [{\citenamefont {van Nieuwenburg}\ \emph {et~al.}(2019)\citenamefont
		{van Nieuwenburg}, \citenamefont {Baum},\ and\ \citenamefont
		{Refael}}]{vanNieuwenburg9269}%
	\BibitemOpen
	\bibfield  {author} {\bibinfo {author} {\bibfnamefont {E.}~\bibnamefont {van
				Nieuwenburg}}, \bibinfo {author} {\bibfnamefont {Y.}~\bibnamefont {Baum}}, \
		and\ \bibinfo {author} {\bibfnamefont {G.}~\bibnamefont {Refael}},\ }\href
	{\doibase 10.1073/pnas.1819316116} {\bibfield  {journal} {\bibinfo  {journal}
			{Proceedings of the National Academy of Sciences}\ }\textbf {\bibinfo
			{volume} {116}},\ \bibinfo {pages} {9269} (\bibinfo {year}
		{2019})}\BibitemShut {NoStop}%
	\bibitem [{\citenamefont {Schulz}\ \emph {et~al.}(2019)\citenamefont {Schulz},
		\citenamefont {Hooley}, \citenamefont {Moessner},\ and\ \citenamefont
		{Pollmann}}]{stark}%
	\BibitemOpen
	\bibfield  {author} {\bibinfo {author} {\bibfnamefont {M.}~\bibnamefont
			{Schulz}}, \bibinfo {author} {\bibfnamefont {C.~A.}\ \bibnamefont {Hooley}},
		\bibinfo {author} {\bibfnamefont {R.}~\bibnamefont {Moessner}}, \ and\
		\bibinfo {author} {\bibfnamefont {F.}~\bibnamefont {Pollmann}},\ }\href
	{\doibase 10.1103/PhysRevLett.122.040606} {\bibfield  {journal} {\bibinfo
			{journal} {Phys. Rev. Lett.}\ }\textbf {\bibinfo {volume} {122}},\ \bibinfo
		{pages} {040606} (\bibinfo {year} {2019})}\BibitemShut {NoStop}%
	\bibitem [{\citenamefont {van Horssen}\ \emph {et~al.}(2015)\citenamefont {van
			Horssen}, \citenamefont {Levi},\ and\ \citenamefont
		{Garrahan}}]{Horssen2015}%
	\BibitemOpen
	\bibfield  {author} {\bibinfo {author} {\bibfnamefont {M.}~\bibnamefont {van
				Horssen}}, \bibinfo {author} {\bibfnamefont {E.}~\bibnamefont {Levi}}, \ and\
		\bibinfo {author} {\bibfnamefont {J.~P.}\ \bibnamefont {Garrahan}},\ }\href
	{\doibase 10.1103/PhysRevB.92.100305} {\bibfield  {journal} {\bibinfo
			{journal} {Phys. Rev. B}\ }\textbf {\bibinfo {volume} {92}},\ \bibinfo
		{pages} {100305} (\bibinfo {year} {2015})}\BibitemShut {NoStop}%
	\bibitem [{\citenamefont {Hickey}\ \emph {et~al.}(2016)\citenamefont {Hickey},
		\citenamefont {Genway},\ and\ \citenamefont {Garrahan}}]{Hickey2016}%
	\BibitemOpen
	\bibfield  {author} {\bibinfo {author} {\bibfnamefont {J.~M.}\ \bibnamefont
			{Hickey}}, \bibinfo {author} {\bibfnamefont {S.}~\bibnamefont {Genway}}, \
		and\ \bibinfo {author} {\bibfnamefont {J.~P.}\ \bibnamefont {Garrahan}},\
	}\href {\doibase 10.1088/1742-5468/2016/05/054047} {\bibfield  {journal}
		{\bibinfo  {journal} {J. Stat. Mech. Theory Exp.}\ }\textbf {\bibinfo
			{volume} {2016}},\ \bibinfo {pages} {054047} (\bibinfo {year}
		{2016})}\BibitemShut {NoStop}%
	\bibitem [{\citenamefont {Lee}\ \emph {et~al.}(2006)\citenamefont {Lee},
		\citenamefont {Nagaosa},\ and\ \citenamefont {Wen}}]{Lee2006doping}%
	\BibitemOpen
	\bibfield  {author} {\bibinfo {author} {\bibfnamefont {P.~A.}\ \bibnamefont
			{Lee}}, \bibinfo {author} {\bibfnamefont {N.}~\bibnamefont {Nagaosa}}, \ and\
		\bibinfo {author} {\bibfnamefont {X.-G.}\ \bibnamefont {Wen}},\ }\href
	{\doibase 10.1103/RevModPhys.78.17} {\bibfield  {journal} {\bibinfo
			{journal} {Rev. Mod. Phys.}\ }\textbf {\bibinfo {volume} {78}},\ \bibinfo
		{pages} {17} (\bibinfo {year} {2006})}\BibitemShut {NoStop}%
	\bibitem [{\citenamefont {Schwinger}(1962)}]{schwinger1962}%
	\BibitemOpen
	\bibfield  {author} {\bibinfo {author} {\bibfnamefont {J.}~\bibnamefont
			{Schwinger}},\ }\href {\doibase 10.1103/PhysRev.128.2425} {\bibfield
		{journal} {\bibinfo  {journal} {Phys. Rev.}\ }\textbf {\bibinfo {volume}
			{128}},\ \bibinfo {pages} {2425} (\bibinfo {year} {1962})}\BibitemShut
	{NoStop}%
	\bibitem [{\citenamefont {Schwinger}(1951)}]{schwinger1951}%
	\BibitemOpen
	\bibfield  {author} {\bibinfo {author} {\bibfnamefont {J.}~\bibnamefont
			{Schwinger}},\ }\href {\doibase 10.1103/PhysRev.82.664} {\bibfield  {journal}
		{\bibinfo  {journal} {Phys. Rev.}\ }\textbf {\bibinfo {volume} {82}},\
		\bibinfo {pages} {664} (\bibinfo {year} {1951})}\BibitemShut {NoStop}%
	\bibitem [{\citenamefont {Hamer}\ \emph {et~al.}(1982)\citenamefont {Hamer},
		\citenamefont {Kogut}, \citenamefont {Crewther},\ and\ \citenamefont
		{Mazzolini}}]{hamer1982massive}%
	\BibitemOpen
	\bibfield  {author} {\bibinfo {author} {\bibfnamefont {C.~J.}\ \bibnamefont
			{Hamer}}, \bibinfo {author} {\bibfnamefont {J.}~\bibnamefont {Kogut}},
		\bibinfo {author} {\bibfnamefont {D.~P.}\ \bibnamefont {Crewther}}, \ and\
		\bibinfo {author} {\bibfnamefont {M.~M.}\ \bibnamefont {Mazzolini}},\ }\href
	{\doibase https://doi.org/10.1016/0550-3213(82)90229-2} {\bibfield  {journal}
		{\bibinfo  {journal} {Nuclear Physics B}\ }\textbf {\bibinfo {volume}
			{208}},\ \bibinfo {pages} {413 } (\bibinfo {year} {1982})}\BibitemShut
	{NoStop}%
	\bibitem [{\citenamefont {Kogut}\ and\ \citenamefont
		{Susskind}(1975)}]{kogut1975j}%
	\BibitemOpen
	\bibfield  {author} {\bibinfo {author} {\bibfnamefont {J.}~\bibnamefont
			{Kogut}}\ and\ \bibinfo {author} {\bibfnamefont {L.}~\bibnamefont
			{Susskind}},\ }\href {\doibase 10.1103/PhysRevD.11.395} {\bibfield  {journal}
		{\bibinfo  {journal} {Phys. Rev. D}\ }\textbf {\bibinfo {volume} {11}},\
		\bibinfo {pages} {395} (\bibinfo {year} {1975})}\BibitemShut {NoStop}%
	\bibitem [{\citenamefont {Nielsen}\ and\ \citenamefont
		{Ninomiya}(1981)}]{nielsen1981absence}%
	\BibitemOpen
	\bibfield  {author} {\bibinfo {author} {\bibfnamefont {H.~B.}\ \bibnamefont
			{Nielsen}}\ and\ \bibinfo {author} {\bibfnamefont {M.}~\bibnamefont
			{Ninomiya}},\ }\href {\doibase https://doi.org/10.1016/0550-3213(81)90361-8}
	{\bibfield  {journal} {\bibinfo  {journal} {Nuclear Physics B}\ }\textbf
		{\bibinfo {volume} {185}},\ \bibinfo {pages} {20 } (\bibinfo {year}
		{1981})}\BibitemShut {NoStop}%
	\bibitem [{\citenamefont {Montvay}\ and\ \citenamefont
		{M{\"u}nster}(1994)}]{montvay1997quantum}%
	\BibitemOpen
	\bibfield  {author} {\bibinfo {author} {\bibfnamefont {I.}~\bibnamefont
			{Montvay}}\ and\ \bibinfo {author} {\bibfnamefont {G.}~\bibnamefont
			{M{\"u}nster}},\ }\href@noop {} {\emph {\bibinfo {title} {Quantum fields on a
				lattice}}}\ (\bibinfo  {publisher} {Cambridge University Press},\ \bibinfo
	{year} {1994})\BibitemShut {NoStop}%
	\bibitem [{\citenamefont {Karsten}\ and\ \citenamefont
		{Smith}(1981)}]{karsten1981lattice}%
	\BibitemOpen
	\bibfield  {author} {\bibinfo {author} {\bibfnamefont {L.~H.}\ \bibnamefont
			{Karsten}}\ and\ \bibinfo {author} {\bibfnamefont {J.}~\bibnamefont
			{Smith}},\ }\href {\doibase https://doi.org/10.1016/0550-3213(81)90549-6}
	{\bibfield  {journal} {\bibinfo  {journal} {Nuclear Physics B}\ }\textbf
		{\bibinfo {volume} {183}},\ \bibinfo {pages} {103 } (\bibinfo {year}
		{1981})}\BibitemShut {NoStop}%
	\bibitem [{\citenamefont {Rothe}(2005)}]{rothe2005lattice}%
	\BibitemOpen
	\bibfield  {author} {\bibinfo {author} {\bibfnamefont {H.~J.}\ \bibnamefont
			{Rothe}},\ }\href@noop {} {\emph {\bibinfo {title} {Lattice Gauge Theories:
				An Introduction}}},\ Vol.~\bibinfo {volume} {74}\ (\bibinfo  {publisher}
	{World Scientific Publishing Company},\ \bibinfo {year} {2005})\BibitemShut
	{NoStop}%
	\bibitem [{\citenamefont {Smith}\ \emph {et~al.}(2018)\citenamefont {Smith},
		\citenamefont {Knolle}, \citenamefont {Moessner},\ and\ \citenamefont
		{Kovrizhin}}]{Smith2018}%
	\BibitemOpen
	\bibfield  {author} {\bibinfo {author} {\bibfnamefont {A.}~\bibnamefont
			{Smith}}, \bibinfo {author} {\bibfnamefont {J.}~\bibnamefont {Knolle}},
		\bibinfo {author} {\bibfnamefont {R.}~\bibnamefont {Moessner}}, \ and\
		\bibinfo {author} {\bibfnamefont {D.~L.}\ \bibnamefont {Kovrizhin}},\ }\href
	{\doibase 10.1103/PhysRevB.97.245137} {\bibfield  {journal} {\bibinfo
			{journal} {Phys. Rev. B}\ }\textbf {\bibinfo {volume} {97}},\ \bibinfo
		{pages} {245137} (\bibinfo {year} {2018})}\BibitemShut {NoStop}%
	\bibitem [{\citenamefont {Hauschild}\ \emph {et~al.}(2016)\citenamefont
		{Hauschild}, \citenamefont {Heidrich-Meisner},\ and\ \citenamefont
		{Pollmann}}]{Hauschild2016}%
	\BibitemOpen
	\bibfield  {author} {\bibinfo {author} {\bibfnamefont {J.}~\bibnamefont
			{Hauschild}}, \bibinfo {author} {\bibfnamefont {F.}~\bibnamefont
			{Heidrich-Meisner}}, \ and\ \bibinfo {author} {\bibfnamefont
			{F.}~\bibnamefont {Pollmann}},\ }\href {\doibase 10.1103/PhysRevB.94.161109}
	{\bibfield  {journal} {\bibinfo  {journal} {Phys. Rev. B}\ }\textbf {\bibinfo
			{volume} {94}},\ \bibinfo {pages} {161109} (\bibinfo {year}
		{2016})}\BibitemShut {NoStop}%
	\bibitem [{\citenamefont {Schneider}\ \emph {et~al.}(2012)\citenamefont
		{Schneider}, \citenamefont {Hackerm{\"u}ller}, \citenamefont {Ronzheimer},
		\citenamefont {Will}, \citenamefont {Braun}, \citenamefont {Best},
		\citenamefont {Bloch}, \citenamefont {Demler}, \citenamefont {Mandt},
		\citenamefont {Rasch} \emph {et~al.}}]{schneider2012fermionic}%
	\BibitemOpen
	\bibfield  {author} {\bibinfo {author} {\bibfnamefont {U.}~\bibnamefont
			{Schneider}}, \bibinfo {author} {\bibfnamefont {L.}~\bibnamefont
			{Hackerm{\"u}ller}}, \bibinfo {author} {\bibfnamefont {J.~P.}\ \bibnamefont
			{Ronzheimer}}, \bibinfo {author} {\bibfnamefont {S.}~\bibnamefont {Will}},
		\bibinfo {author} {\bibfnamefont {S.}~\bibnamefont {Braun}}, \bibinfo
		{author} {\bibfnamefont {T.}~\bibnamefont {Best}}, \bibinfo {author}
		{\bibfnamefont {I.}~\bibnamefont {Bloch}}, \bibinfo {author} {\bibfnamefont
			{E.}~\bibnamefont {Demler}}, \bibinfo {author} {\bibfnamefont
			{S.}~\bibnamefont {Mandt}}, \bibinfo {author} {\bibfnamefont
			{D.}~\bibnamefont {Rasch}},  \emph {et~al.},\ }\href@noop {} {\bibfield
		{journal} {\bibinfo  {journal} {Nature Physics}\ }\textbf {\bibinfo {volume}
			{8}},\ \bibinfo {pages} {213} (\bibinfo {year} {2012})}\BibitemShut {NoStop}%
	\bibitem [{\citenamefont {Choi}\ \emph
		{et~al.}(2016{\natexlab{a}})\citenamefont {Choi}, \citenamefont {Hild},
		\citenamefont {Zeiher}, \citenamefont {Schau{\ss}}, \citenamefont
		{Rubio-Abadal}, \citenamefont {Yefsah}, \citenamefont {Khemani},
		\citenamefont {Huse}, \citenamefont {Bloch},\ and\ \citenamefont
		{Gross}}]{Choi1547}%
	\BibitemOpen
	\bibfield  {author} {\bibinfo {author} {\bibfnamefont {J.-y.}\ \bibnamefont
			{Choi}}, \bibinfo {author} {\bibfnamefont {S.}~\bibnamefont {Hild}}, \bibinfo
		{author} {\bibfnamefont {J.}~\bibnamefont {Zeiher}}, \bibinfo {author}
		{\bibfnamefont {P.}~\bibnamefont {Schau{\ss}}}, \bibinfo {author}
		{\bibfnamefont {A.}~\bibnamefont {Rubio-Abadal}}, \bibinfo {author}
		{\bibfnamefont {T.}~\bibnamefont {Yefsah}}, \bibinfo {author} {\bibfnamefont
			{V.}~\bibnamefont {Khemani}}, \bibinfo {author} {\bibfnamefont {D.~A.}\
			\bibnamefont {Huse}}, \bibinfo {author} {\bibfnamefont {I.}~\bibnamefont
			{Bloch}}, \ and\ \bibinfo {author} {\bibfnamefont {C.}~\bibnamefont
			{Gross}},\ }\href {\doibase 10.1126/science.aaf8834} {\bibfield  {journal}
		{\bibinfo  {journal} {Science}\ }\textbf {\bibinfo {volume} {352}},\ \bibinfo
		{pages} {1547} (\bibinfo {year} {2016}{\natexlab{a}})}\BibitemShut {NoStop}%
	\bibitem [{\citenamefont {Susskind}(1977)}]{susskind1977lattice}%
	\BibitemOpen
	\bibfield  {author} {\bibinfo {author} {\bibfnamefont {L.}~\bibnamefont
			{Susskind}},\ }\href {\doibase 10.1103/PhysRevD.16.3031} {\bibfield
		{journal} {\bibinfo  {journal} {Phys. Rev. D}\ }\textbf {\bibinfo {volume}
			{16}},\ \bibinfo {pages} {3031} (\bibinfo {year} {1977})}\BibitemShut
	{NoStop}%
	\bibitem [{\citenamefont {Hamer}\ \emph {et~al.}(1997)\citenamefont {Hamer},
		\citenamefont {Weihong},\ and\ \citenamefont {Oitmaa}}]{Hamer_1997}%
	\BibitemOpen
	\bibfield  {author} {\bibinfo {author} {\bibfnamefont {C.~J.}\ \bibnamefont
			{Hamer}}, \bibinfo {author} {\bibfnamefont {Z.}~\bibnamefont {Weihong}}, \
		and\ \bibinfo {author} {\bibfnamefont {J.}~\bibnamefont {Oitmaa}},\ }\href
	{\doibase 10.1103/physrevd.56.55} {\bibfield  {journal} {\bibinfo  {journal}
			{Physical Review D}\ }\textbf {\bibinfo {volume} {56}},\ \bibinfo {pages}
		{55} (\bibinfo {year} {1997})}\BibitemShut {NoStop}%
	\bibitem [{\citenamefont {Peskin}\ and\ \citenamefont
		{Schroeder}(1995)}]{peskin1995introduction}%
	\BibitemOpen
	\bibfield  {author} {\bibinfo {author} {\bibfnamefont {M.~E.}\ \bibnamefont
			{Peskin}}\ and\ \bibinfo {author} {\bibfnamefont {D.~V.}\ \bibnamefont
			{Schroeder}},\ }\href {http://www.slac.stanford.edu/~mpeskin/QFT.html} {\emph
		{\bibinfo {title} {{An Introduction to quantum field theory}}}}\ (\bibinfo
	{publisher} {Addison-Wesley},\ \bibinfo {address} {Reading, USA},\ \bibinfo
	{year} {1995})\BibitemShut {NoStop}%
	\bibitem [{\citenamefont {Wilson}(1974)}]{wilson1974confinement}%
	\BibitemOpen
	\bibfield  {author} {\bibinfo {author} {\bibfnamefont {K.~G.}\ \bibnamefont
			{Wilson}},\ }\href@noop {} {\bibfield  {journal} {\bibinfo  {journal}
			{Physical review D}\ }\textbf {\bibinfo {volume} {10}},\ \bibinfo {pages}
		{2445} (\bibinfo {year} {1974})}\BibitemShut {NoStop}%
	\bibitem [{\citenamefont {Wilson}(1977)}]{wilson1977new}%
	\BibitemOpen
	\bibfield  {author} {\bibinfo {author} {\bibfnamefont {K.~G.}\ \bibnamefont
			{Wilson}},\ }\href@noop {} {\enquote {\bibinfo {title} {New phenomena in
				subnuclear physics},}\ } (\bibinfo {year} {1977})\BibitemShut {NoStop}%
	\bibitem [{\citenamefont {Chandrasekharan}\ and\ \citenamefont
		{Wiese}(1997)}]{chandrasekharan1997quantum}%
	\BibitemOpen
	\bibfield  {author} {\bibinfo {author} {\bibfnamefont {S.}~\bibnamefont
			{Chandrasekharan}}\ and\ \bibinfo {author} {\bibfnamefont {U.-J.}\
			\bibnamefont {Wiese}},\ }\href@noop {} {\bibfield  {journal} {\bibinfo
			{journal} {Nuclear Physics B}\ }\textbf {\bibinfo {volume} {492}},\ \bibinfo
		{pages} {455} (\bibinfo {year} {1997})}\BibitemShut {NoStop}%
	\bibitem [{\citenamefont {Sala}\ \emph {et~al.}(2018)\citenamefont {Sala},
		\citenamefont {Shi}, \citenamefont {K{\"{u}}hn}, \citenamefont
		{Ba{\~{n}}uls}, \citenamefont {Demler},\ and\ \citenamefont
		{Cirac}}]{Sala2018}%
	\BibitemOpen
	\bibfield  {author} {\bibinfo {author} {\bibfnamefont {P.}~\bibnamefont
			{Sala}}, \bibinfo {author} {\bibfnamefont {T.}~\bibnamefont {Shi}}, \bibinfo
		{author} {\bibfnamefont {‚.~S.}\ \bibnamefont {K{\"{u}}hn}}, \bibinfo
		{author} {\bibfnamefont {M.~C.}\ \bibnamefont {Ba{\~{n}}uls}}, \bibinfo
		{author} {\bibfnamefont {E.}~\bibnamefont {Demler}}, \ and\ \bibinfo {author}
		{\bibfnamefont {J.~I.}\ \bibnamefont {Cirac}},\ }\href {\doibase
		10.1103/PhysRevD.98.034505} {\bibfield  {journal} {\bibinfo  {journal} {Phys.
				Rev. D}\ }\textbf {\bibinfo {volume} {98}} (\bibinfo {year} {2018}),\
		10.1103/PhysRevD.98.034505}\BibitemShut {NoStop}%
	\bibitem [{\citenamefont {Choi}\ \emph
		{et~al.}(2016{\natexlab{b}})\citenamefont {Choi}, \citenamefont {Hild},
		\citenamefont {Zeiher}, \citenamefont {Schauss}, \citenamefont
		{Rubio-Abadal}, \citenamefont {Yefsah}, \citenamefont {Khemani},
		\citenamefont {Huse}, \citenamefont {Bloch},\ and\ \citenamefont
		{Gross}}]{Choi2016}%
	\BibitemOpen
	\bibfield  {author} {\bibinfo {author} {\bibfnamefont {J.-y.}\ \bibnamefont
			{Choi}}, \bibinfo {author} {\bibfnamefont {S.}~\bibnamefont {Hild}}, \bibinfo
		{author} {\bibfnamefont {J.}~\bibnamefont {Zeiher}}, \bibinfo {author}
		{\bibfnamefont {P.}~\bibnamefont {Schauss}}, \bibinfo {author} {\bibfnamefont
			{A.}~\bibnamefont {Rubio-Abadal}}, \bibinfo {author} {\bibfnamefont
			{T.}~\bibnamefont {Yefsah}}, \bibinfo {author} {\bibfnamefont
			{V.}~\bibnamefont {Khemani}}, \bibinfo {author} {\bibfnamefont {D.~A.}\
			\bibnamefont {Huse}}, \bibinfo {author} {\bibfnamefont {I.}~\bibnamefont
			{Bloch}}, \ and\ \bibinfo {author} {\bibfnamefont {C.}~\bibnamefont
			{Gross}},\ }\href {\doibase 10.1126/science.aaf8834} {\bibfield  {journal}
		{\bibinfo  {journal} {Science (80-. ).}\ }\textbf {\bibinfo {volume} {352}},\
		\bibinfo {pages} {1547} (\bibinfo {year} {2016}{\natexlab{b}})}\BibitemShut
	{NoStop}%
	\bibitem [{\citenamefont {Schreiber}\ \emph {et~al.}(2015)\citenamefont
		{Schreiber}, \citenamefont {Hodgman}, \citenamefont {Bordia}, \citenamefont
		{Luschen}, \citenamefont {Fischer}, \citenamefont {Vosk}, \citenamefont
		{Altman}, \citenamefont {Schneider},\ and\ \citenamefont
		{Bloch}}]{Schreiber2015}%
	\BibitemOpen
	\bibfield  {author} {\bibinfo {author} {\bibfnamefont {M.}~\bibnamefont
			{Schreiber}}, \bibinfo {author} {\bibfnamefont {S.~S.}\ \bibnamefont
			{Hodgman}}, \bibinfo {author} {\bibfnamefont {P.}~\bibnamefont {Bordia}},
		\bibinfo {author} {\bibfnamefont {H.~P.}\ \bibnamefont {Luschen}}, \bibinfo
		{author} {\bibfnamefont {M.~H.}\ \bibnamefont {Fischer}}, \bibinfo {author}
		{\bibfnamefont {R.}~\bibnamefont {Vosk}}, \bibinfo {author} {\bibfnamefont
			{E.}~\bibnamefont {Altman}}, \bibinfo {author} {\bibfnamefont
			{U.}~\bibnamefont {Schneider}}, \ and\ \bibinfo {author} {\bibfnamefont
			{I.}~\bibnamefont {Bloch}},\ }\href {\doibase 10.1126/science.aaa7432}
	{\bibfield  {journal} {\bibinfo  {journal} {Science (80-. ).}\ }\textbf
		{\bibinfo {volume} {349}},\ \bibinfo {pages} {842} (\bibinfo {year}
		{2015})}\BibitemShut {NoStop}%
	\bibitem [{\citenamefont {Smith}(2019)}]{smith2019localization}%
	\BibitemOpen
	\bibfield  {author} {\bibinfo {author} {\bibfnamefont {A.}~\bibnamefont
			{Smith}},\ }in\ \href@noop {} {\emph {\bibinfo {booktitle} {Disorder-Free
				Localization}}}\ (\bibinfo  {publisher} {Springer},\ \bibinfo {year} {2019})\
	pp.\ \bibinfo {pages} {55--69}\BibitemShut {NoStop}%
	\bibitem [{\citenamefont {Zache}\ \emph {et~al.}(2018)\citenamefont {Zache},
		\citenamefont {Hebenstreit}, \citenamefont {Jendrzejewski}, \citenamefont
		{Oberthaler}, \citenamefont {Berges},\ and\ \citenamefont
		{Hauke}}]{zache2018quantum}%
	\BibitemOpen
	\bibfield  {author} {\bibinfo {author} {\bibfnamefont {T.~V.}\ \bibnamefont
			{Zache}}, \bibinfo {author} {\bibfnamefont {F.}~\bibnamefont {Hebenstreit}},
		\bibinfo {author} {\bibfnamefont {F.}~\bibnamefont {Jendrzejewski}}, \bibinfo
		{author} {\bibfnamefont {M.~K.}\ \bibnamefont {Oberthaler}}, \bibinfo
		{author} {\bibfnamefont {J.}~\bibnamefont {Berges}}, \ and\ \bibinfo {author}
		{\bibfnamefont {P.}~\bibnamefont {Hauke}},\ }\href {\doibase
		10.1088/2058-9565/aac33b} {\bibfield  {journal} {\bibinfo  {journal} {Quantum
				Science and Technology}\ }\textbf {\bibinfo {volume} {3}},\ \bibinfo {pages}
		{034010} (\bibinfo {year} {2018})}\BibitemShut {NoStop}%
\end{thebibliography}
\end{document}